\documentclass[aps,showpacs,tighten,preprint,
prl]{revtex4}
\usepackage{graphicx}

\begin{document}

\title{Coherent Resonances Observed in the Dissociative Electron Attachments to Carbon
Monoxide}

\author{Xu-Dong Wang\dag, Chuan-Jin Xuan\dag, Yi Luo\dag\ddag,}
\author{Shan Xi Tian\dag}
   \email{sxtian@ustc.edu.cn}
\affiliation{\dag Hefei National Laboratory for Physical Sciences at
the Microscale, Center of Advanced Chemical Physics and Department
of Chemical Physics, University of Science and Technology of China,
Hefei, Anhui 230026, China} \affiliation{\ddag Department of
Theoretical Chemistry and Biology, School of Biotechnology, Royal
Institute of Technology, S-106 91, Stockholm, Sweden}

\begin{abstract}
Succeeding our previous finding about coherent interference of the
resonant states of CO$^-$ formed by the low-energy electron
attachment [Phys. Rev. A {\bf 88}, 012708 (2013)], here we provide
more evidences of the coherent interference, in particular, we find
the state configuration change in the interference with the increase
of electron attachment energy by measuring the completely backward
distributions of the O$^-$ fragment ion of the temporary CO$^-$ in
an energy range 11.3 - 12.6 eV. Therefore, different pure states,
namely, coherent resonances, can be formed when the close-lying
resonant states are coherently superposed by a broad-band electron
pulse.

\end{abstract} 

\maketitle

Dissociative electron attachment (DEA) is a novel electron inelastic
scattering process, in which the temporary negative ion is a
resonant system ($e^-$-M) for an atomic or molecular target M
\cite{taylor}. This temporary anion M$^-$ will undergo dissociation
or electron auto-detachment \cite{christo}. By measuring the
momentum distributions of the negatively charged fragment, one can
have insights into the formation and dissociation dynamics of M$^-$
\cite{boud, hall, konig, tian}. Different resonant states of M$^-$
can be formed in the low-energy electron attachments to M, and
locate energetically lower or higher than the eigenstates of the
neutral M, due to the coupling between the the eigenstate and the
continuum background of the excess $e^-$ \cite{christo, wang}.
Obviously, the resonant states easily interact or couple with each
other by the aid of the continuum background when these states are
energetically close \cite{moise}. In our recent study of the DEA to
carbon monoxide molecule, a coherent resonance as the result of
quantum interference among $^2\Pi$, $^2\Delta$, and $^2\Phi$
resonant states of CO$^-$ was found at 10.6 eV and responsible for
the completely backward distribution of the fast O$^-$ fragment ion
\cite{tian}. The fast O$^-$ ion is produced via $e^-$ + CO
$\rightarrow$ C($^3$P)+O$^-$ (Process I), while the much slow one
(with the kinetic energy near zero eV, the left panel of FIG. 2c in
ref.\cite{tian}) should be the yield of Process II [$e^-$ + CO
$\rightarrow$ C($^1$D)+O$^-$] \cite{hall}. With the increase of the
attachment energy, the slow O$^-$ ion becomes fast by sharing the
excess energy with the metastable C($^1$D). It is unknown whether
the same or similar quantum interference influences the momentum
distributions of the O$^-$ ion from Process II.

In this communication, we report that the {\it coherent resonance}
continuously leads to the rainbow-like momentum distributions of the
O$^-$ ion produced via Process II, namely, the backward angular
distribution with respect to the electron incident direction, by
using anion velocity time-sliced map imaging
technique\cite{tian,wu}. More interestingly, with the increase of
attachment energy, the coherent interference is changed from the
configuration of $^2\Pi$, $^2\Delta$, and $^2\Phi$ resonant states
to that of $^2\Sigma$, $^2\Delta$, and $^2\Phi$ resonant states,
implying the existences of two different coherent resonances. A new
dissociation channel, $e^-$ + CO $\rightarrow$ C($^1$S)+O$^-$
(Process III), is found in the measurement at 12.3 eV, which was not
reported before \cite{hall}.

The experiments were done with our home-made anion momentum imaging
apparatus \cite{tian,wu}. In brief, an effusive molecular beam of CO
(along {\it y} axis) was perpendicular to the pulsed electron beam
(along {\it x} axis, the thermal energy spread of 0.5 eV) which was
collimated with the homogeneous magnetic field produced by a pair of
Helmholtz coils. The O$^-$ ions were periodically pushed out of the
reaction area (volume size was less than 2$\times$2$\times$2 mm$^3$)
and expanded to form a Newton sphere which passed through a series
of the circle electrodes and time-of-flight tube (along {\it y}
axis). The high momentum resolution ($\triangle v/v \leq$ 2$\%$) was
achieved by these well-designed electrodes \cite{wu}. The O$^-$ ions
were detected with a set of triple microchannel plates and a
phosphor screen (its diameter is 40 mm). The central time-sliced
image of O$^-$ was recorded by the application of a narrow time-gate
pulse voltage (width 45 ns) on the last microchanel plate and with a
CCD camera. In our previous experiments \cite{tian}, the Newton
sphere of the O$^-$ ions produced via Process II was too small to be
sliced. According to the working principle of the ion velocity slice
imaging technique, this Newton sphere was magnified by reducing the
voltages on the pusher and the other electrodes in a proper ratio
\cite{li}. The kinetic energy of the ion fragment was obtained after
calibration with the experimental data available in the literature.

As shown in FIG. 1(a-e), the O$^-$ momentum images for Process II
appear to grow in size radially with the increase of the electron
attachment energy; while the most of the ions produced in Process I
are lost (FIG. 1a and 1b) because the Newton sphere of the latter is
too large and out of the detector. At the electron attachment
energies of 12.4 and 12.6 eV, the much slower O$^-$ ions appear as
the central points of the images in FIG. 1(d) and 1(e). These slower
ions are produced in Process III, because the threshold energy of
this process is 12.30 eV \cite{cal}. The kinetic energies of the
O$^-$ ions via Process II are plotted against the electron
attachment energy in FIG. 1(f). The fitted line intersects the {\it
x} axis at 10.9 eV, in excellent agreement with the calculated
threshold 10.88 eV of Process II \cite{cal}. Our present and
previous \cite{xia} studies show the high experimental accuracy in
determining the dissociation thresholds. More importantly, the
angular distributions of the O$^-$ ions produced in both Processes
II and III clearly indicate the backward scattering character, which
is quite similar to that observed in Process I \cite{tian}.

As discussed in our previous study for Process I, the coherent
interference of $^2\Pi$, $^2\Delta$, and $^2\Phi$ resonant states
was proposed to interpret the completely backward distribution of
O$^-$ at 10.6 eV \cite{tian}. Is this mechanism still applicable for
the present observations? The O$^-$ angular distributions of Process
II are plotted in FIG. 2 where the ion intensities are obtained by
integrations of the ion signals within the selected kinetic energy
ranges. On the basis of the theoretical model for the fragment
momentum distribution in the DEA to diatomic molecule
\cite{omalley}, the differential cross section $\sigma_{DEA}$ is,
\begin{equation}
\sigma_{DEA}\propto
\sum_{|\mu|}|\sum_{j=1,l=|\mu|}c_jie^{i\delta_j}Y_{l\mu}(\theta,\zeta)|^2
\end{equation}
where the incident electron is described with a series of partial
waves with different angular momenta $l$, $c_j$ is the weighing
parameters of each wave $j$, $\delta_j$ is the phase lag among the
partial waves due to the interaction between the incoming electron
and target, and $Y_{l\mu}$ is the spherical harmonics. The parameter
$\mu$ has a absolute value of $|\Lambda_f-\Lambda_i|$, representing
the difference in the projection of the angular momenta along the
internuclear axis for the neutral ($i$) and anionic resonant ($f$)
states. In the simple cases, one or several resonant states (mixed
state) should be considered, corresponding to one or serval $|\mu|$
values. If two or more resonant states $\varphi_i$ as a
superposition state, $\varphi =\sum \varphi_i, i= 1, 2, ...$,
contribute to the cross section, the quantum interference among
these resonant states $\varphi_i$ is introduced in a straightforward
way \cite{tian},
\begin{equation}
\sigma_{DEA}\propto
\sum_{\alpha,\beta}I_{\alpha,\beta}+2\sum_{\alpha\neq\beta}\sqrt{I_{\alpha}I_{\beta}}\cos\phi_{\alpha\beta}
\end{equation}
where $I_{\alpha}$ is the amplitude contribution in the form of
$|\sum c_jie^{i\delta_j}Y_{l\mu}|^2$ for resonant state $\alpha$ and
$\phi_{\alpha\beta}$ is the phase shift or difference between
$\alpha$ and $\beta$ states. Following the procedure proposed in our
previous work \cite{tian}, one or several states (eq.1) and then the
interference form (eq.2) were used in fitting the experimental
angular distributions, indicating the best fitting results by using
eq.2 (the solid and dashed curves in FIG. 2).

The solid curves represent the data fitting by considering of the
quantum interference of $^2\Pi$, $^2\Delta$, and $^2\Phi$ resonant
states, where two partial waves are used in each $I$ term: $l =
1(p)$ and $2(d)$ for $^2\Pi$, $l = 2(d)$ and $3(f)$ for $^2\Delta$,
and $l = 3(f)$ and $4(g)$ for $^2\Phi$ \cite{tian}. At the energies
higher than 11.8 eV, two configurations of the state interference,
$^2\Pi + ^2\Delta + ^2\Phi$ and $^2\Sigma + ^2\Delta + ^2\Phi$, are
used, where $^2\Sigma$ is the fifth resonant state of CO$^-$ and all
of them are close-lying in the Franck-Condon region of the electron
vertical attachment \cite{morgan}. The interference of $^2\Pi$,
$^2\Delta$, and $^2\Phi$ resonant states can give the good fittings
at the lower energies 11.3 and 11.8 eV, while some bumps appear and
the fitting curves clearly deviate from the experimental data in the
forward direction at the higher energies 12.4 and 12.6 eV. The
interference of $^2\Sigma$, $^2\Delta$, and $^2\Phi$ resonant states
can describe no distributions in the forward direction (see the
dashed curves in FIG. 2) at these higher energies. At 12.1 eV, we
cannot identify the best one from these two interference modes. In
Table I, the fitting parameters are listed. We notice that the sum
of phase shifts among the resonant states is consistently satisfied
with $\phi_{12}+\phi_{23}+\phi_{31} \approx 2\pi$ rad. The phase sum
of $2\pi$ evidences the coherency of quantum interference or the
existence of coherent resonance; moreover, two coherence resonances
are observed in this work.

To reveal the different contributions of all terms in eq.2, the
angular distribution from each term is plotted in FIG. 3 by using
the fitting parameters. In FIG. 3a, the sum of all terms is the
fitting curve at 11.3 eV for the interference of $^2\Pi$,
$^2\Delta$, and $^2\Phi$ states. The $^2\Phi$ state component shows
the maximum around 80$^{\circ}$, two maxima appear at 60$^{\circ}$
and 120$^{\circ}$ for $^2\Delta$ state, while one maximum around
130$^{\circ}$ arises from $^2\Pi$ state. The forward distributions
from $\Delta$, $\Phi$, and the cross term $\Pi * \Delta$ are
counteracted with the cross terms $\Delta *\Phi$ and $\Pi * \Phi$.
In FIG. 3b, the sum curve corresponds to the fitting result at 12.6
eV for the interference of $^2\Sigma$, $^2\Delta$, and $^2\Phi$
states. The $^2\Sigma$ state typically shows both the forward and
backward distributions, and the forward ones are overall canceled
with the cross terms $\Sigma *\Phi$ and $\Delta * \Phi$. In general,
the cross terms, i.e., the interference terms
$I_{\alpha}I_{\beta}\cos\phi_{\alpha\beta}$, play the essential
roles in the completely backward distributions of the O$^-$ ions.

Although the statistic errors of the ion signal are large for
Process III, the angular distributions at 12.6 eV in FIG. 4 also
indicate the backward scattering character. The peak positions of
the ion intensities show some differences when the O$^-$ ions are
selected in the different kinetic energy ranges. Considering the
uncertain weighings of the ions with the lower kinetic energies, we
only fit the angular distribution of the ions within the higher
kinetic energy range of 0.15-0.20 eV by using eq.2 with the
interference of $^2\Sigma$, $^2\Delta$, $^2\Phi$ states, indicating
a good fitting (dashed) curve in FIG. 4. As listed in Table I, the
phase sum $\phi_{12}+\phi_{23}+\phi_{31}$ approximately equals
2$\pi$ rad again, implying that the corresponding coherence
resonance should also be responsible for the backward distributions
of the O$^-$ ion produced via Process III.

In summary, we demonstrate that the coherent interference of
$^2\Pi$, $^2\Delta$, and $^2\Phi$ resonant states of CO$^-$ can be
established in a wide energy range, but the state configuration of
the interference is changed to  $^2\Sigma$, $^2\Delta$, and $^2\Phi$
resonant states when the electron attachment energy is higher than
12.1 eV. All of the sums of phase shifts obtained in the
experimental data fittings are equal to 2$\pi$ rad, indicating that
two coherent resonances exist as the superposition sates, $\varphi_1
= ^2\Pi +^2\Delta + ^2\Phi$ and $\varphi_2 = ^2\Sigma +^2\Delta +
^2\Phi$. We believe that the formation of the coherent resonance is
similar to the production of coherent wave packet using a broad-band
ultrashort laser pulse in coherent control \cite{ohmori}, but it is
a new concept in quantum scattering and deserves the further
theoretical explorations.

This work is supported by the NSFC (Grant No. 21273213) and MOST
(Grant No. 2013CB834602).

\newpage

Table I. Fitting parameters obtained by using eq.2 for the angular
distributions of O$^-$ produced via Processes II and III.

\begin{center}
\begin{tabular}{lllll}
  \hline
    $^2\Pi +^2\Delta + ^2\Phi$ & & & \\
    \hline
Attachment Energy (eV)  & 11.3 (P II) & 11.8 (P II) & 12.1 (P II)\\
 Weighing ratio & & & \\
$c_1 : c_2: c_3:$ &0.98:0.52:1.00: &0.90:0.46:0.37: & 0.68:0.49:1.00:\\
$ c_4: c_5: c_6$ &0.99:1.67:0.31 &0.68:1.00:0.62 & 0.46:0.52:0.25\\
Phase lags (rad) & & & \\
($^2\Pi$)$\delta_d -\delta_p$ & 1.57&1.57 & 0.40\\
($^2\Delta$)$\delta_f -\delta_d$ &3.05 &-1.36 & -1.57\\
($^2\Phi$)$\delta_g -\delta_f$ &-1.56 &-1.77 & -2.28\\
Phase shift (rad) & & & \\
$\phi_{12}$($^2\Pi-^2\Delta$) &0.00 &0.16 & 3.10\\
$\phi_{23}$($^2\Delta-^2\Pi$) &3.01 &3.20 & 0.00\\
$\phi_{31}$($^2\Phi-^2\Pi$) & 3.07 &3.15 & 3.16\\
\hline
  $^2\Sigma +^2\Delta + ^2\Phi$  &  & & \\
\hline
Attachment Energy (eV) & 12.4 (P II) & 12.6 (P II) & 12.6 (P III)\\
 Weighing ratio & & & \\
$c_1 : c_2: c_3:$ & 0.05:0.12:1.00: & 0.10:0.44:1.00: & 0.11:0.19:1.00:\\
$c_4: c_5: c_6$ &0.41:0.80:0.00 & 0.78:0.67:0.32 & 0.26:0.97:0.04\\
Phase lags (rad) & & & \\
($^2\Sigma$)$\delta_p -\delta_s$ & 0.40&0.21 & 1.91\\
($^2\Delta$)$\delta_f -\delta_d$ & 2.35&0.04 & 1.04\\
($^2\Phi$)$\delta_g -\delta_f$ &-3.14 &-1.57 & 1.69\\
Phase shift (rad) & & & \\
$\phi_{12}$($^2\Sigma-^2\Delta$) &1.84 &1.43 & 1.53\\
$\phi_{23}$($^2\Delta-^2\Sigma$) &1.19 &2.33 & 1.39\\
$\phi_{31}$($^2\Phi-^2\Sigma$) & 3.29 &2.46 &3.14\\
  \hline
\end{tabular}
\end{center}

\newpage
\begin{center}
\includegraphics[width=160mm]{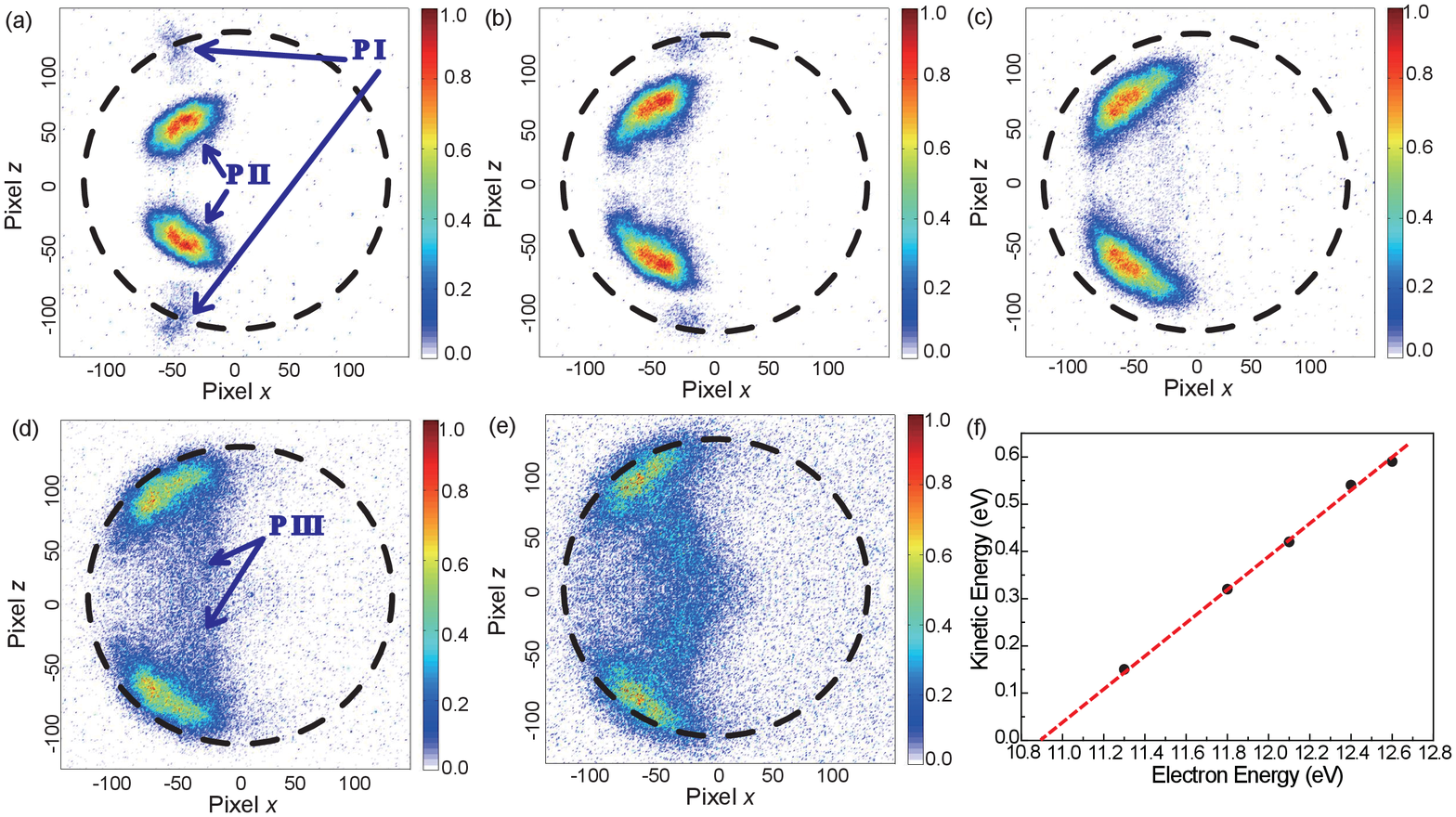}
\end{center}

FIG. 1: Sliced images of O$^-$ recorded at 11.3 (a), 11.8 (b), 12.1
(c), 12.4(d), and 12.6 eV (e), where the electron incident direction
is from left to right and through the image center, the ion
intensity is normalized respectively, P I, P II, P III denote the
different dissociation processes (see the details in text), and the
broken circle represents the effective area (diameter 40 mmm) of the
detector. (f) The most probable kinetic energy of O$^-$ (produced
via P II) in terms of the electron attachment energy.

\newpage

\begin{center}
\includegraphics[width=100mm]{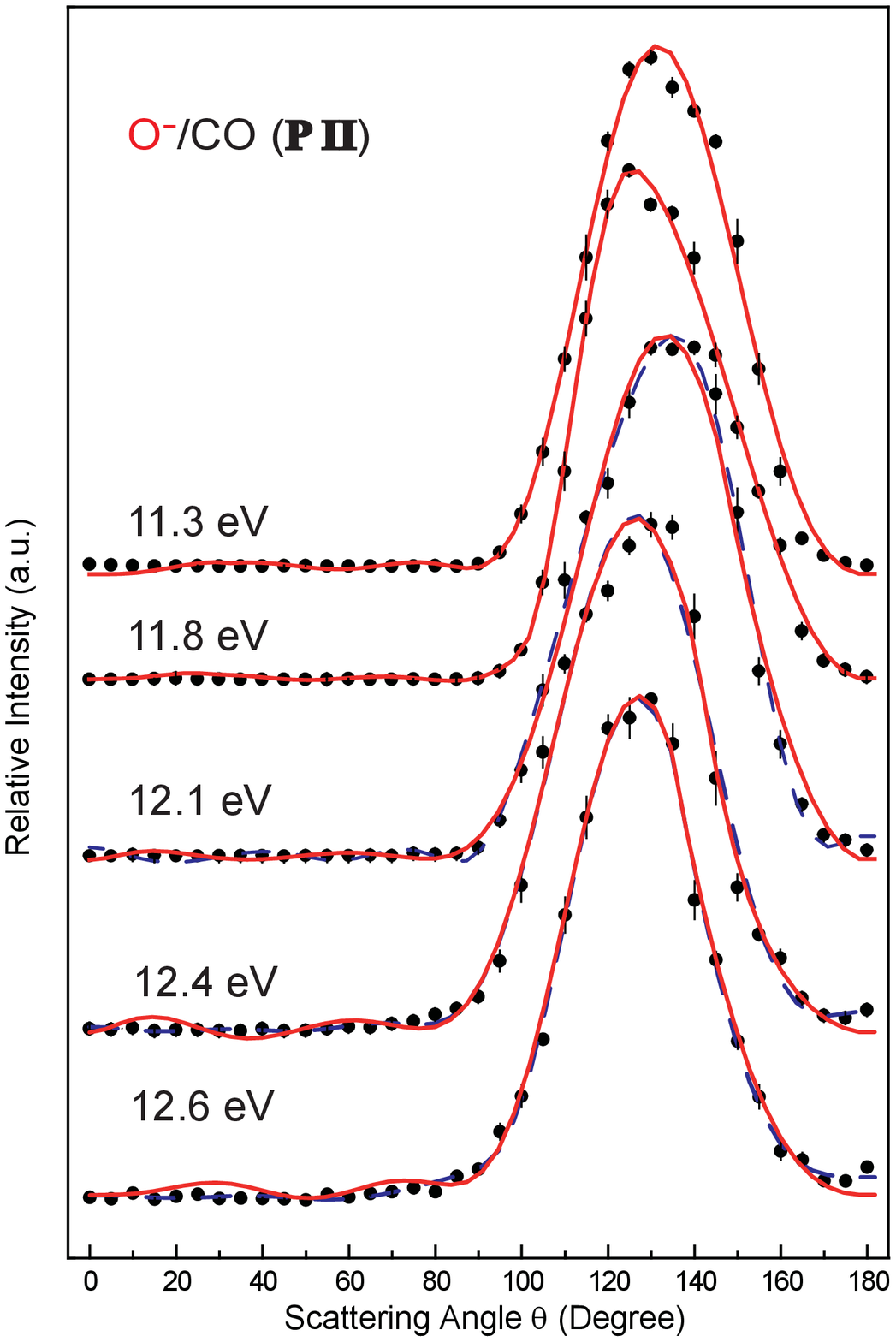}
\end{center}

FIG. 2 Angular distribution of O$^-$ ions from Process II: solid
circles, the experimental data (the selected kinetic energy range:
0.14-0.24 eV at 11.3 eV, 0.25-0.35 eV at 11.8 eV, 0.36-0.46 eV at
12.1 eV, 0.49-0.59 eV at 12.4 eV, 0.54-0.64 eV at 12.6 eV); solid
curves, the data fitting with the quantum interference of $^2\Pi$,
$^2\Delta$, and $^2\Phi$; dashed curves, the data fitting with the
quantum interference of $^2\Sigma$, $^2\Delta$, and $^2\Phi$.

\newpage

\begin{center}
\includegraphics[width=100mm]{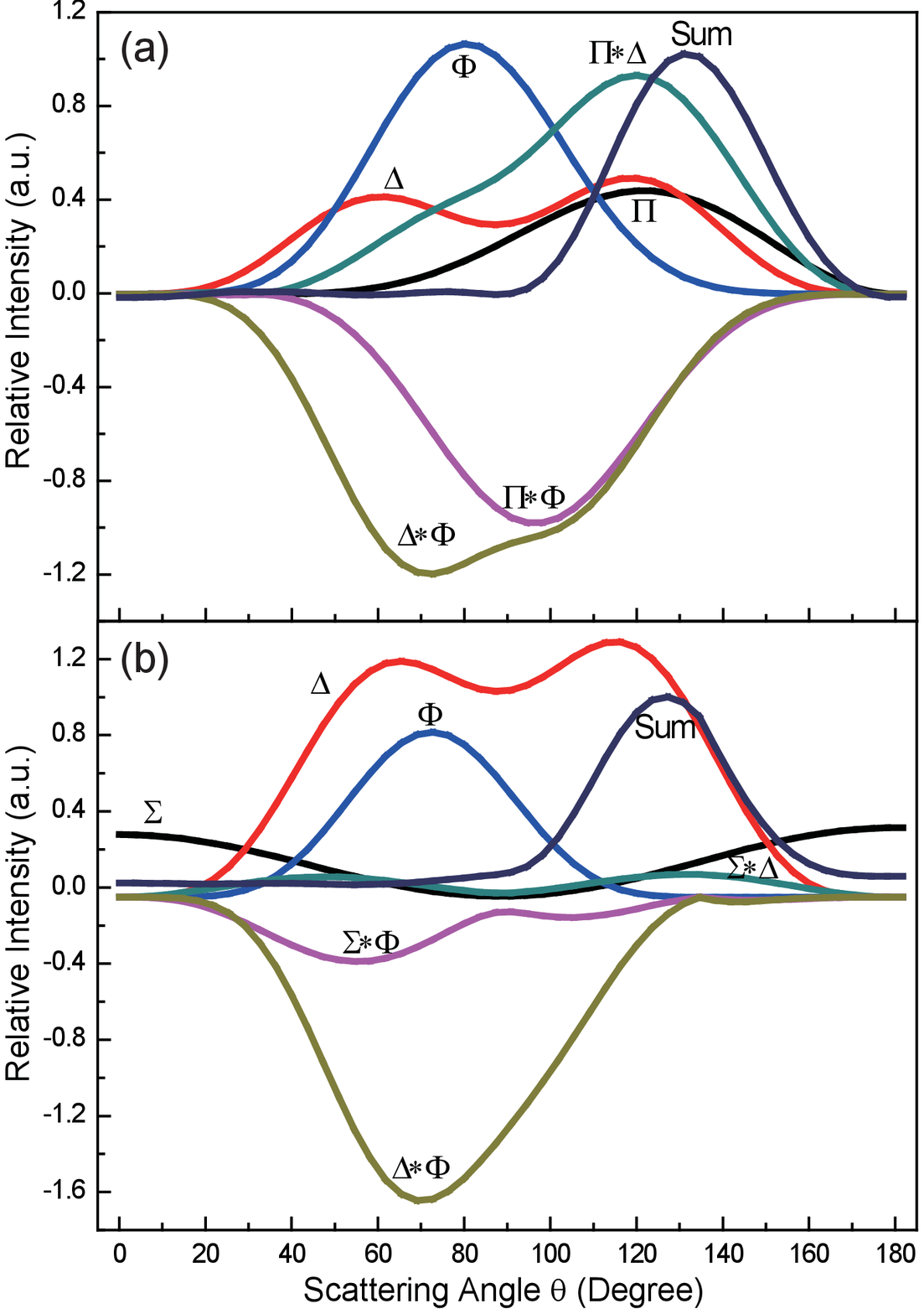}
\end{center}

FIG. 3 Contributions of each term in eq.2 to the angular
distribution using the fitting parameters for the quantum
interference of $^2\Pi$, $^2\Delta$, and $^2\Phi$ at 11.3 eV (a) and
that of $^2\Sigma$, $^2\Delta$, and $^2\Phi$ at 12.6 eV (b).

\newpage

\begin{center}
\includegraphics[width=100mm]{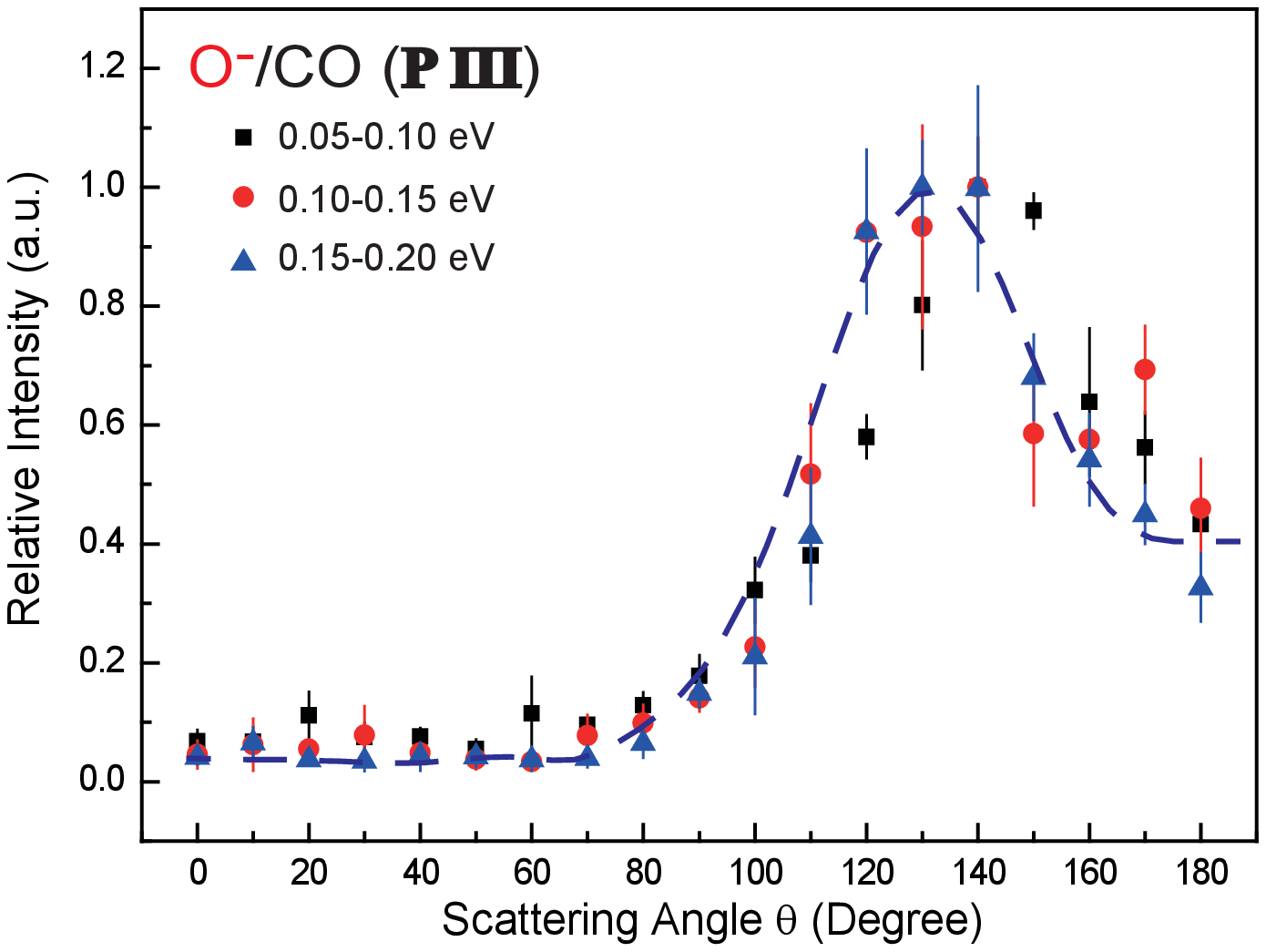}
\end{center}

FIG. 4 Angular distribution of O$^-$ ions from Process III at 12.6
eV where the dashed curve represents the data fitting with the
quantum interference of $^2\Sigma$, $^2\Delta$, and $^2\Phi$ for the
ions with the kinetic energy of 0.15-0.20 eV.


\begin{thebibliography}{}

\bibitem{taylor} J. R. Taylor, \emph{Scattering Theory: The Quantum Theory of Norelativistic Collisions},
(John Wiely \& Sons, New York, 1972).

\bibitem{christo} \emph{Electron-Molecule Interactions and Their Applications}, edited by L. G.
Christophorou (Academic Press, New York, 1984).

\bibitem{boud} B. Bouda\"{\i}fa, P. Cloutier, D. Hunting, M. A. Huels, and L. Sanche, Science \textbf{287}, 1658 (2000).

\bibitem{hall} R. I. Hall, I. \v{C}ade\v{z}, C. Schermann, and M.
Tronc, Phys. Rev. A \textbf{15}, 599 (1977).

\bibitem{konig} C. K\"{o}nig, J. Kopyra, I. Bald, and E.
Illenberger, Phys. Rev. Lett. \textbf{97}, 018105 (2006).

\bibitem{tian} S. X. Tian, B. Wu, L. Xia, Y.-F. Wang, H.-K. Li,
X.-J. Zeng, Y. Luo, and J. Yang, Phys. Rev. A \textbf{88}, 012708
(2013).

\bibitem{wang} Y.-F. Wang and S. X. Tian, Phys. Rev. A \textbf{85},
052709 (2012).

\bibitem{moise} E. Narevicius and N. Moiseyev, Phys. Rev. Lett.
\textbf{84}, 1681 (2000).

\bibitem{wu} B. Wu, L. Xia, H.-K. Li, X.-J. Zeng, and S. X. Tian,
Rev. Sci. Instrum. \textbf{83}, 013108 (2012).

\bibitem{li} H.-K. Li, L. Xia, Z.-J. Zeng, and S. X. Tian, J. Phys.
Chem. A \textbf{117}, 3176 (2013).

\bibitem{cal} The threshold of Process I is 9.62 eV \cite{hall}, and C($^1$D) and C($^1$S) are 1.26 and 2.68 eV higher than
C($^3$P), respectively [V. Kaufman and J. F. Ward, J. Opt. Soc. Am.
\textbf{56}, 1591 (1966)].

\bibitem{xia} L. Xia, X.-D. Wang, C.-J. Xuan, X.-J. Zeng, H.-K. Li, S. X.
Tian, Y. Pan, and K.-C. Lau, J. Chem. Phys. \textbf{140}, 041106
(2014).

\bibitem{omalley} T. F. O'Malley, Phys. Rev. \textbf{150}, 14
(1966); T. F. O'Malley and H. S. Taylor, {\it ibid}. \textbf{176},
207 (1968).

\bibitem{morgan} L. A. Morgan and J. Tennyson, J. Phys. B
\textbf{26}, 2429 (1993).

\bibitem{ohmori} K. Ohmori, Annu. Rev. Phys. Chem. \textbf{60}, 487
(2009).

\end{thebibliography}
\end{document}